# Possible sublimation and dust activity on primitive NEAs: Example of (162173) Ryugu

Vladimir V. Busarev[a, b,] *, Faith Vilas[c], Andrei B. Makalkin[d]

[a]*Lomonosov Moscow State University, Sternberg Astronomical Institute, University Av., 13, Moscow, 119992, Russian Federation (RF)*

[b]*Institute of Astronomy, Russian Academy of Sciences, Pyatnitskaya St. 48, 109017 Moscow, RF*

[c]*Planetary Science Institute, 1700 E. Fort Lowell Rd., Suite 106, Tucson, Arizona, 85719, USA*

[d]*Schmidt Institute of Physics of the Earth, Russian Academy of Sciences, Bolshaya Gruzinskaya str., 10-1, Moscow, 123242, RF*

* Corresponding author.

E-mail address: busarev@sai.msu.ru (V.V. Busarev)

## Abstract

Because of the relatively high surface temperatures and the rapid evolution of orbital parameters of near-Earth asteroids (NEAs) of primitive types, the possibility of retention of icy materials (predominately $H_2O$ ice) is questionable. Yet, based on consideration of all ground-based spectroscopic observations of Japanese *Hayabusa 2* space mission target (162173) Ryugu (Cg-type) obtained to date, we suspected transient sublimation activity occuring on the asteroid. However, unlike some main-belt primitive asteroids demonstrating the sublimation of ices close to their perihelion distances and, respectively, at the highest subsolar temperatures (Busarev et al., 2015, 2017a, 2017b), the effect on Ryugu was likely observed in an unexpected location – at the passage of aphelion. To explain the difference, we calculated the subsolar temperature depending on the NEA's heliocentric distance and performed some analytical estimations related to internal structure and thermo-physical parameters of such bodies.

The presumed temporal sublimation/dust activity on Ryugu, as an asteroid of primitive type, could be an indication of a cracked monolithic structure and existence of a residual frozen core and/or closed gas traps that could point to a relatively recent transition of the asteroid from the main asteroid belt to the near-Earth region. As the calculations show, the discussed phenomenon on Ryugu could point to a borderline case between primitive bodies of greater and lesser size, placed in the same conditions where the larger bodies can maintain an ice core, and the smaller bodies would keep only gaseous volatiles.

## 1. Introduction

The abundance of volatile compounds on near-Earth asteroids (NEAs) of primitive types is still unknown. It is assumed that such asteroids are minor solid celestial bodies with predominantly low-temperature mineralogy or Tholen's C, F, B, G, D, and P types (Tholen, 1989). As in the case of an extinct cometary nucleus, volatiles could be hidden in the NEAs' interiors, under a desiccated crust. Difficulties with ground-based observations of these bodies (due to their small size, high proper motion, quick change of orbital parameters, etc.) make the detection of probable transient phenomena connected with the sublimation of their volatiles (predominantly $H_2O$ ice), and removal of dust particles, almost impossible. Only the delivery of a primitive NEA's sample to the Earth by spacecraft would clearly answer this question. Two such missions are already underway: Japan's *Hayabusa 2* space mission, which will return a sample of (162173) Ryugu (Cg-type) in 2020 (e. g., Abe et al., 2007; Ishiguro et al., 2014), and NASA's *OSIRIS-REx* mission which will study (101955) Bennu (B-type) and return a sample in 2023 (e. g., Clark et al., 2011).

Importantly, because of the proximity of NEAs to the Earth, they are thought to be the parent bodies of the majority of meteorites in terrestrial collections (e. g., Gaffey et al., 1993; Binzel et al., 2001; McSween et al., 2006). In general, the composition, density and structure of meteorites of different chemical groups can be used to infer characteristics of asteroids of different taxonomic types. Here porosity is a key parameter. Ordinary chondrites typically have porosity of under 10%, while most carbonaceous chondrites are more than 20% porous, and CI meteorites can be up to 35% porous (e. g., Consolmagno et al., 2008). As argued by Consolmagno et al (2008) in their review of meteorite density and porosity, when meteoritic densities are "compared with the densities of small solar system bodies, one can deduce the nature of asteroid and comet interiors, which in turn reflect the accretional and collisional conditions in the early solar system". Similarly, microporosity of chondrites, and especially of carbonaceous chondrites, probably indicates the previous presence of frozen volatiles in the matter of primitive asteroids.

We analyzed the visible-range spectral reflectance features of some main-belt primitive asteroids demonstrating sublimation activity perihelion (Busarev et al., 2015, 2016, 2017a, 2017b), all published results of spectroscopic observations of (162173) Ryugu, and took into account our analytical estimations of thermophysical characteristics of primitive NEAs based on suitable parameters of carbonaceous chondrites and similar compound (e. g., Wechsler et al., 1972; Frost et al., 2000; Britt et al., 2002; Opeil et al., 2010, 2012;). The data have led us to a suggestion about the possibility of some periodic transient sublimation activity of the asteroid.

## 2. Description of obtained reflectance spectra of Ryugu and other data

The main parameters of (162173) Ryugu, a potentially hazardous Apollo NEA, are as follows: diameter ~0.87 km (Ishiguro et al., 2014) and geometric albedo in the range 0.044 to 0.050 (Müller et al., 2017) with average value of 0.047; Cg-class asteroid (Binzel et al., 2001, 2002); and orbital and rotation periods of 1.3 yr (http://ssd.jpl.nasa.gov/sbdb.cgi#top) and 7.63109 h (Müller et al., 2017), respectively. As dynamical modeling shows, the asteroid might have been be delivered to near-Earth space from the inner main-belt region (2.15 AU < $a$ < 2.5 AU, $i$ < 8º) due to the action of the $v_6$ resonance (Campins et al., 2013).

We have analyzed all Ryugu's reflectance spectra obtained by different authors to date. Examples of such spectra from the publications of Binzel et al. (2001), Vilas (2008, 2012) and Lazzaro et al. (2013) are shown in Figs. 1 - 3. We use the relative rotational phase (RRP = 0) of Ryugu counted from the reflectance spectrum of 1999 (Binzel et al., 2001) to characterize the orientation of the asteroid and compare the reflectance spectra on a common scale taking into account the most reliable rotational period of the asteroid $T_{rot.}$ = 7.63109 h calculated from a combined analysis of optical and radiometric data (Müller et al., 2017). The values of the RRP along with other parameters (if available) are collected in Table 1. They are listed in chronological order and briefly described. As it turned out, each of the other sets of Ryugu's reflectance spectra obtained by other observers were acquired at relatively close RRP – those by Vilas (2008), Moskovitz et al. (2013), Lazzaro et al. (2013) (see Table 1), and (Sugita et al., 2013). Observational data from Sugita et al. (2013) are not included in Table 1 because the exact time (UT) of the observations is absent. We consider these data as confirmations of the rotational period value, $T_{rot.}$ = 7.63109 h (Müller et al., 2017), and our calculations of RRPs corresponding to separate reflectance spectra.

The RRP value marked in bold lettering (Table 1) designate the closest one in RRP (0.109) to Ryugu's most discussed reflectance spectrum on 11 July 2007 (with unusual maximum at ~0.56 μm, RRP = 0.139) by Vilas (2008). Bold values of heliocentric distance ($r$) correspond to reflectance spectra of the asteroid with increased dispersion of points as an indication of onset of hypothetical sublimation activity.

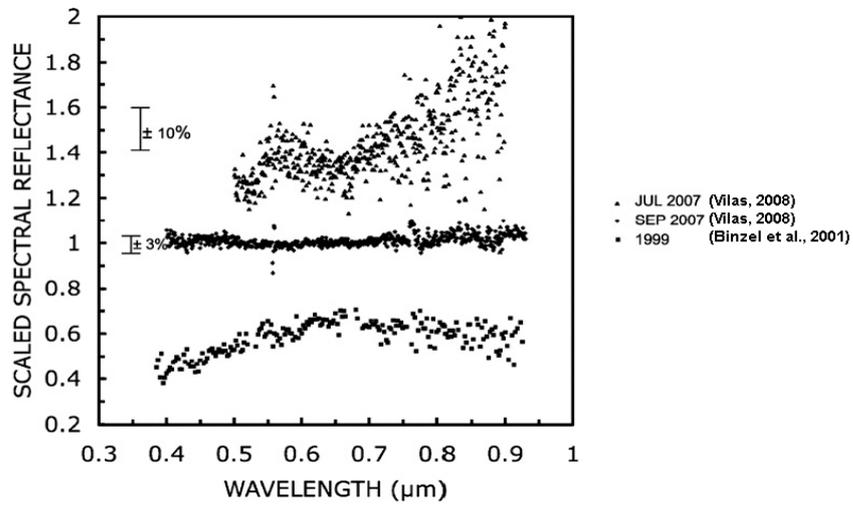

Fig. 1. Normalized at 0.55 μm and offset reflectance spectra of (162173) Ryugu reproduced from article by Vilas (2008). Two upper spectra were obtained on 2007 July 11 and 2007 September 10/11 (composite), respectively. The lowermost spectrum of Ryugu was obtained during its 1999 discovery apparition by Binzel et al. (2001). Error bars represent scattering of points in the spectra.

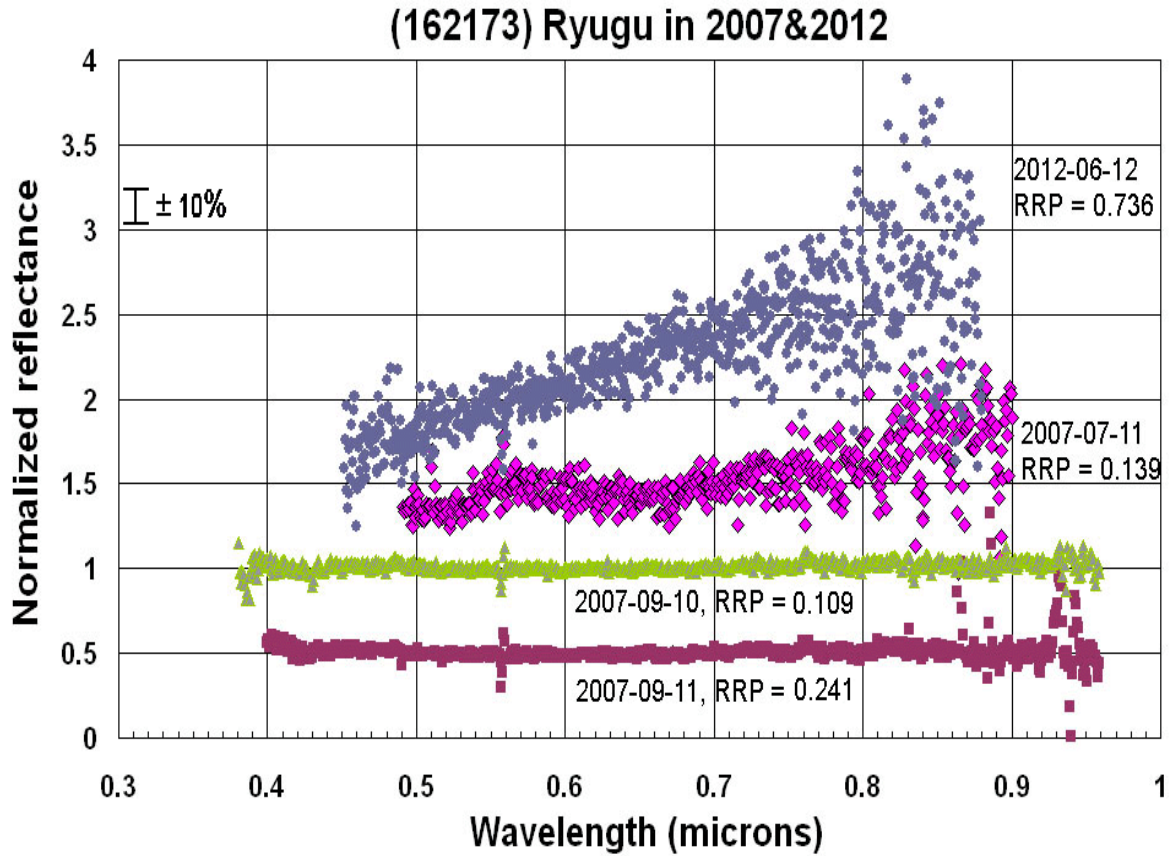

Fig. 2. All reflectance spectra of (162173) Ryugu (normalized to unit value at 0.55 μm, and offset by intervals of 0.5 for clarity) obtained by Vilas (2008, 2012). The relative rotational phases (RRP) of the asteroid are counted from the average observation time of its reflectance spectrum by Binzel et al. (2001) in 1999 (assuming the RRP = 0). The scale of relative errors is shown in the top left corner of the figure.

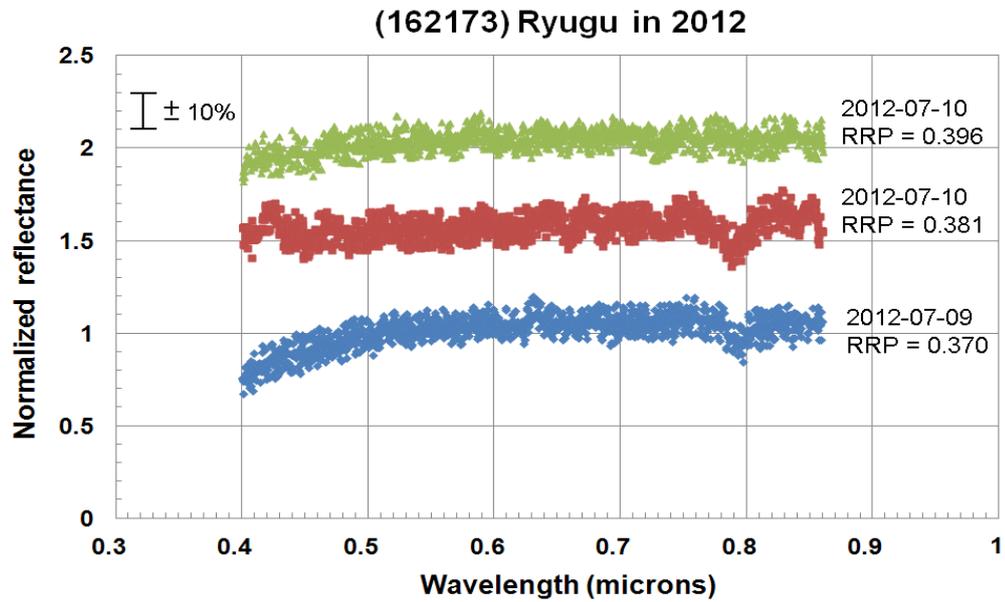

Fig. 3. As in Fig. 1 for reflectance spectra of (162173) Ryugu obtained by Lazzaro et al. (2013).

Table 1. Observational parameters of (162173) Ryugu at the times (the average time of exposure) of its spectroscopic observations in chronological order ($T_{rot.}$ = 7.6309 h; q = 0.9632060 AU, Q = 1.4159152 AU).

| Dates | UT (h:m:s), | r (AU) | Before or after aphelion | Ph (°) | V | RRP | Description of the spectrum |
|---|---|---|---|---|---|---|---|
| 1999 05 17 [a] | 07:49:00 | 1.314 | b/a | 6.1 | 17.8 | **0** | convex noisy spectrum with a maximum at 0.65 μm |
| 2007 07 11 [b] | 10:34:44 | **1.381** | a/a | 40.3 | 20.4 | **0.139** | very noisy spectrum with a maximum at 0.57 μm (Fig. 1) |
| 2007 09 10 [b] | 06:11:23 | 1.254 | a/a | 22.4 | 18.0 | **0.109** | the best in quality, smooth and flat spectrum |
| 2007 09 11 [b] | 06:13:50 | 1.251 | a/a | 22.8 | 18.0 | 0.241 | smooth and flat spectrum |
| 2012 06 01 [c] | 07:07:00 | 1.367 | b/a | 0.2 | 17.8 | 0.300 | smooth and flat reflectance spectrum (spectrum with designation "A" in original publication) |
| 2012 06 01 [c] | 08:09:00 | 1.367 | b/a | 0.2 | 17.8 | 0.306 | smooth and flat reflectance spectrum (spectrum with designation "B" in original publication) |
| 2012 06 02 [c] | 03:54:00 | 1.368 | b/a | 1.0 | 17.9 | 0.413 | flat spectrum with increase of noise (spectrum with designations "C" in original publication) |
| 2012 06 03 [c] | 01:54:00 | **1.370** | b/a | 2.0 | 18.0 | 0.534 | noisy flat spectra (spectra with designations "D" in original publication |
| 2012 06 03 [c] | 03:34:00 | **1.370** | b/a | 2.1 | 18.0 | 0.543 | noisy flat spectra (spectra with designations "E"in original publication |
| 2012 06 03 [c] | 04:54:00 | **1.370** | b/a | 2.2 | 18.0 | 0.550 | noisy flat spectra (spectra with designations "F" in original publication |
| 2012 06 12 [d] | 06:06:14 | **1.383** | b/a | 12.0 | 18.6 | 0.736 | very noisy spectrum with considerable positive gradient (Fig. 1) |
| 2012 07 09 [e] | 23:35:00 | **1.410** | b/a | 33.0 | 20.0 | 0.370 | convex (in the short-wavelength range) noisy spectrum (Fig 2) |
| 2012 07 10 [e] | 01:37:00 | **1.410** | b/a | 33.1 | 20.0 | 0.381 | flat noisy spectrum (Fig 2) |
| 2012 07 10 [e] | 04:21:00 | **1.410** | b/a | 33.1 | 20.0 | 0.396 | flat noisy spectrum (Fig 2) |

Description of designations and notes in the table: UT – universal time of observations (average); r – heliocentric distance of the asteroid; Ph – phase angle of the asteroid; V – visual stellar magnitude of the asteroid; RRP – relative rotational phase of the asteroid; [a] Binzel et al. (2001); [b] Vilas (2008); [c] Moskovitz et al. (2013); [d] Vilas (2012); [e] Lazzaro et al. (2013); the RRP value marked in bold lettering (Table 1) designate the closest one in RRP (0.109) to Ryugu's most discussed reflectance spectrum on 11 July 2007 (with unusual maximum at ~0.56 μm, RRP

= 0.139) by Vilas (2008). Bold values of heliocentric distance (*r*) correspond to reflectance spectra of the asteroid with increased dispersion of points as an indication of onset of hypothetical sublimation activity.

As seen from Figs. 1-3 and Table 1, all reflectance spectra of Ryugu, except for those by Vilas (2008, 2012), are similar in shape. However, three reflectance spectra of Ryugu by Vilas (2008, 2012) (Fig. 2) corresponding to close distances to aphelion of the asteroid show strong changes: the two on 11 July 2007 (with an unusual maximum at ~0.56 μm, RRP = 0.139) and on 12 June 2012 are very different in shape and noisy, but the spectrum on 10 September 2007 (RRP = 0.109) is flat and of best quality despite of its most proximity in RRP with previous one (Fig. 2). We interpret the strong changes and a rise of scatter (dispersion of points) in reflectance of the asteroid as manifestations of a temporal non-uniform coma or jets connected with a transient $H_2O$ ice sublimation on the body.

Some attempts were made to explain the unusual maximum at ~0.56 μm in the spectrum of Ryugu on 11 July 2007 as an absorption band of phyllosicates at 0.7 μm (Vilas, 2008; Sugita et al., 2013). Such a band is known in reflectance spectra of many primitive asteroids arising due to $Fe^{2+} \rightarrow Fe^{3+}$ charge transfer (e. g., Vilas et al., 1994). However, as follows from consideration of all obtained spectral data on the asteroid to date at different rotational phases, there are not any signs of such band. If it presents in the spectrum of Ryugu (RRP = 0.139), it must be seen in another reflectance spectrum of the asteroid obtained at closest rotational phase, RRP = 0.109 (Fig. 2) (Vilas, 2008).

Additionally we found a similarity (a considerable rise of short-wavelength reflectance) between Ryugu's reflectance spectrum on 11 July 2007 (Vilas, 2008) and spectra of the three main-belt primitive asteroids (145) Adeona, (704) Interamnia, and (779) Nina that showed signs of simultaneous sublimation activity near perihelion (Busarev et al., 2015, 2016, 2017a, 2017b). The same phenomenon was likely detected on (1) Ceres by different methods (A'Hearn, Feldman, 1992; Küppers et al., 2014; Nathues et al., 2015). The most intense sublimation activity on Ceres was found by the *Dawn* spacecraft (NASA) around bright spots (being probable ice formations) at the bottom of some large craters (Nathues et al., 2015). For comparison, normalized reflectance spectra of Ryugu, Adeona, Interamnia, Nina, and Ceres with suggested signs of sublimation activity (maxima in the range 0.48 – 0.56 μm) are shown in Figure 4. Notably, the reflectance spectrum of Ryugu on 11 July 2007 is very scattered (which could be produced by a scattering medium between the asteroid and observer) with a maximum at ~0.56 μm. We do not know so far all conditions of such maximum emerging. It may be an instant feature of a heterogeneous envelope. Another interesting feature of the spectrum is that it was observed nearly a month after Ryugu's passage past aphelion instead of near perihelion. This is unusual, because cometary-like activity of a minor planet is expected near perihelion due to higher temperatures. To explain the difference, we discuss possible mechanisms of ice preservation on this and similar primitive NEAs.

A subtle cloud or coma of sublimed micron-sized particles surrounding an asteroid could produce considerable scattering of light reflected from the asteroid. Given the spectrally neutral refractive index of water ice and "dry ice" particles in the spectral range used here, the wavelength position of a maximum in an asteroid reflectance spectrum is likely determined by a predominant particle size in the scattering coma. Actually, as follows from Mie theory and subsequent numeric modeling (Mie, 1908; Hansen and Travis, 1974), for particles with refractive index of water ice ($n = 1.33$), the maximum scattering intensity is governed by the dimensionless parameter $x = (2\pi \cdot \rho / \lambda) \approx 6$ or $\rho / \lambda \approx 1$, where $\rho$ is the radius of the scattering particles. The average radius of the scattering particles ($\rho$) and spectral position of the maximum of the scattered light ($\lambda$) are matching in this case. We estimate the value of the wavelength of the scattered light for Ryugu (Fig. 4, top spectrum) as ~0.56 μm, which is close to similar values of main-belt asteroids having signs of sublimation activity, and is intermediate between the values for the most primitive (145) Adeona and Mars-crosser (1474) Beira (with more likely dust activity) (Busarev et al., 2016). Similar to the discussed main-belt asteroids, Ryugu's spectrum on 2007 July 11 (Vilas, 2008) has signs of scattering light reflected from the surface in the asteroid's coma formed by sublimed ice and possibly dust particles. However, the spectrum corresponds to a time after aphelion passage by the asteroid, when the subsolar temperature was close to minimum (Fig. 5).

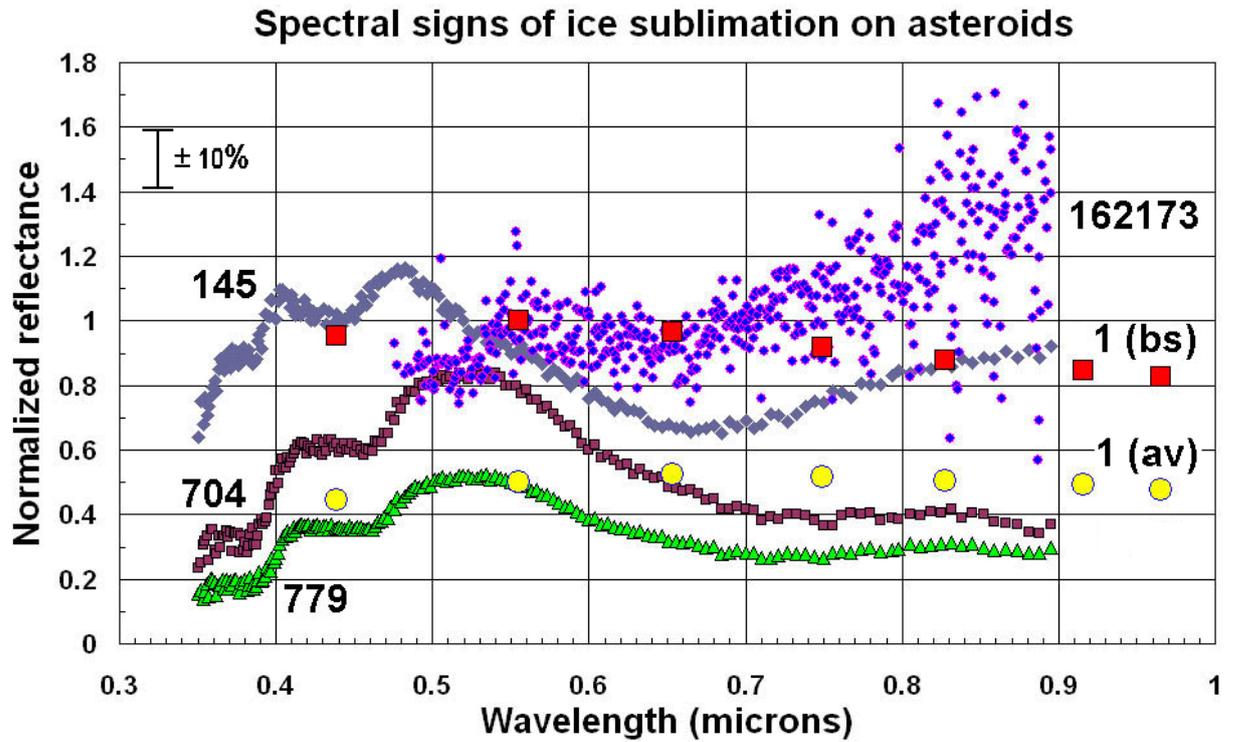

Fig. 4. Reflectance spectra (normalized to unit value at 0.55 μm, and offset by 0.1, 0.2 or 0.5 for clarity) of (162173) Ryugu (on 2007-07-11), (145) Adeona, (704) Interamnia, (779) Nina, and (1) Ceres) showing detected signs of sublimation activity (reflectance maxima in the spectral range 0.48 – 0.55 μm suggesting). Averaged spectra of asteroids 145, 704, and 779 are from Busarev et al. (2015). Spectra of (1) Ceres (Nathues et al., 2015): 1 (bs) – averaged spectrum of bright spots, 1 (av) – averaged spectrum of Ceres's dark surface. The scale of relative errors is shown in the top left corner of the figure.

**3. Estimations of thermophysical parameters and possible mechanisms of sublimation activity on NEAs**

Let us imagine a relatively small primitive-type asteroid recently arrived to the region of terrestrial planets from the main asteroid belt. As follows from the thermophysical model calculation, its surface and internal matter could include some icy component (e. g., Schorghofer, 2016). When a main-belt primitive asteroid becomes a NEA due to a strong impact event or the action of gravitational resonances, it quickly (after several revolutions around the Sun) loses most of ice stored near the surface due to considerable growth of its subsolar and average subsurface temperatures by ~100 degrees.

We estimated changes in the subsolar temperature on/under the surface of observed primitive asteroids. Taking into account the values of Stefan-Boltzmann and solar constants (assuming asteroids emit thermal radiation nearly as black bodies) and the radiative energy received by the unit area normal to the direction to the Sun at heliocentric distance $r$, we obtained a simple formula for estimation of instantaneous effective temperature in the subsolar point ($T_{ss}$) on the equator of an atmosphereless solid body with zero thermal inertia, located at heliocentric distance $r$ (in AU) (Busarev et al., 2015):

$$T_{ss} = 394 \text{ K} \cdot ((1-p_v)/r^2)^{1/4}, \qquad (1)$$

where $p_v$ is the geometric albedo. Calculations using this formula give changes of the subsolar temperature on an asteroid with time (and heliocentric distance) including moments of the observations.

There might be several sources of hypothetical activity on a primitive asteroid. Possible sources of water ice could be in its interiors or in cold traps on its surface near the poles. For implementation of the first possibility on a NEA, the ice must survive in interiors of the asteroid during its residence in near-Earth orbit. However, tidal interactions with the Earth and other terrestrial planets could increase the number and sizes of fractures in the asteroid body that facilitate both venting of gaseous volatiles from its subsurface layers and their loss from the uppermost regolith. A similar effect could be produced (but only in the subsurface layers) by a sharp temperature difference between day and night sides of a NEA with rotation timescale of several hours.

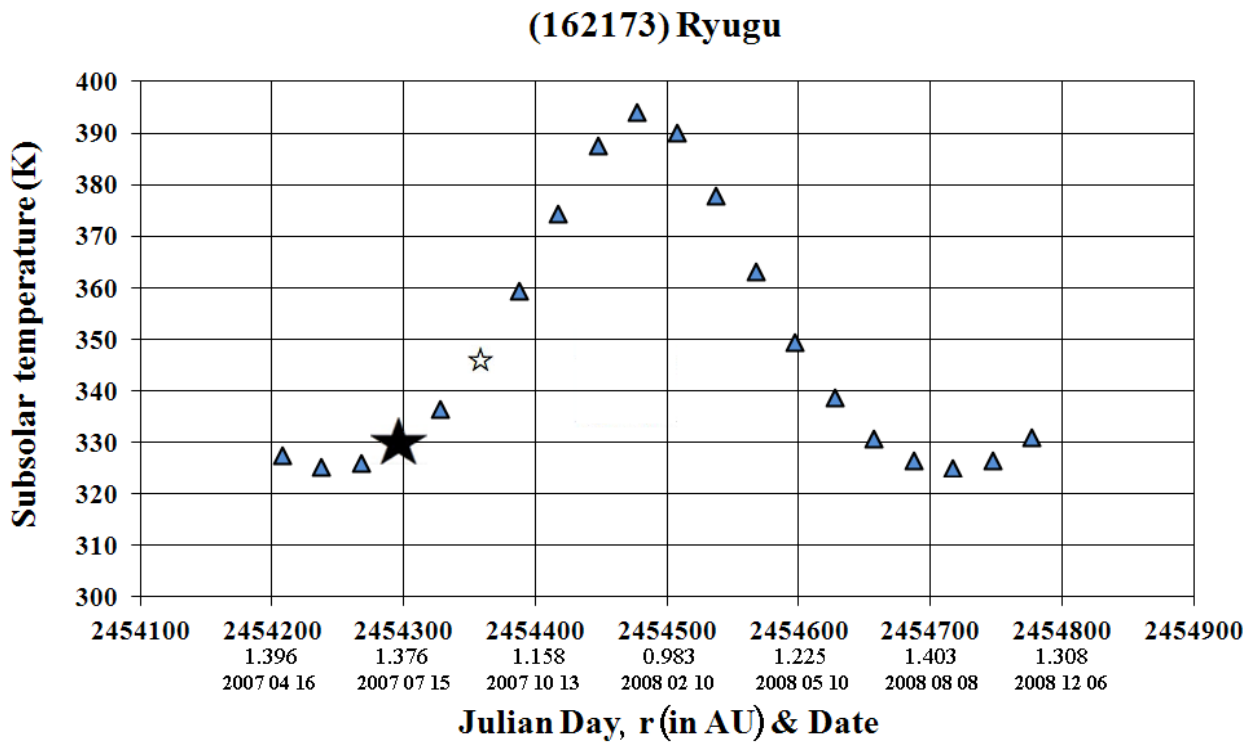

Fig. 5. The subsolar (equatorial) temperature on (162173) Ryugu while orbiting the Sun. Values of temperature and heliocentric distance of the asteroid at the time of observations (Vilas,

2008) are marked by stars: the large one matches 2007 July 11, the instant of possible sublimation activity, and small (open) one matches 2007 September 10/11.

As NEAs, heated by solar radiation, rotate and move in elliptic orbits, the variable insolation generates temperature oscillations in the diurnal and seasonal skin layers. The latter is much thicker than the former, but does not exceed a few meters, as follows from our calculations presented below. Under the seasonal layer, the temperature in asteroid interiors should be constant or decrease with depth. The former option takes place, if the asteroid has moved to the near-Earth orbit from the asteroid belt sufficiently long ago to have time to warm up. The second option corresponds to the case when the asteroid did not have enough time to lose its ice and warm up. The constant temperature in the NEA's interiors can be calculated by the equation for equilibrium temperature of a celestial body heated by the sun and reradiating with emissivity ε, close to 0.9 (Lissauer, de Pater, 2013)

$$T_{eq} = \left( \frac{S_0}{r_{AU}^2} \frac{(1 - A_B)}{4\varepsilon\sigma} \right)^{1/4}, \qquad (2)$$

where $S_0$=1365 W m$^{-2}$ is the solar constant, $r_{AU}^2$ is the heliocentric distance of the body, expressed in AU, $A_B$ is the Bond albedo of the body, σ is the Stefan-Boltzmann constant. For estimation of $T_{eq}$ for Ryugu, we take $r_{AU}$ equal to the semi-major axis of its orbit $a$=1.189 AU (http://ssd.jpl.nasa.gov/sbdb.cgi#top). The Bond albedo can be represented as the product of the geometrical albedo $p_v$ and the phase integral $q$ (e. g., Li et al, 2015), that is, $A_B$= $p_v$ $q$. The geometric V-band albedo for Ryugu, $p_v$=0.049±0.006 (Müller et al., 2017), is very close to that of another NEA, (101955) Bennu ($p_v$=0.047±0.011) (Yu, Ji, 2015). For Ryugu, there are no data on its phase integral $q$. But because Bennu and Ryugu belong to similar taxonomic classes (B and Cg, respectively) and could have close photometric properties, the value obtained for Bennu, $q$ = 0.235 (Yu, Ji, 2015), may be used as a fist approximation for Ryugu. To estimate the equilibrium temperature of Ryugu according to Eq. (2), we use the above parameter values (including the emissivity ε=0.9) and obtain $T_{eq}$=261 K. Actually, the exact value of phase integral $q$ is not of high significance: even if we use in the above estimation the geometrical albedo instead of the Bond albedo the equilibrium temperature will be only 1% lower, that is, 258 K. If we account for slower motion of the asteroid near aphelion than near perihelion, $T_{eq}$ would be a few kelvins lower than the above value, but not lower than 250 K.

It is worthwhile comparing this temperature estimate with the surface temperatures averaged over the orbital period of the asteroid. The calculated subsolar temperature on the surface of Ryugu ranges from ~326- to 394 K (Fig. 5). The subsolar temperature on the surface of Ryugu, averaged over its orbital period, is ~ 350 K taking into account slower motion of the asteroid near aphelion than at perihelion. This estimate is in agreement with a reference maximum (subsolar) temperature given by Müller et al. (2017). So we can take ~350 K as the mean noon temperature on the Ryugu's surface and ~150 K as the mean midnight surface temperature according to thermophysical modeling for a similar object, (101955) Bennu (Yu, Ji, 2015). The average of the two above values, 250 K, could be considered as Ryugu's mean temperature. This value is only a few degrees lower than the equilibrium temperature, $T_{eq}$, estimated above with the help of Eq. (2).

These results show that if Ryugu is on its present orbit for sufficiently long time, its internal temperature would reach the equilibrium value of ~250–260 K.

It is worthwhile comparing the estimated equilibrium temperature with the condensation-evaporation temperature in the center of Ryugu. The pressure $p$ in the centre of an asteroid with density $\rho$ independent on depth is defined by equation

$$p = (\pi/6) G \rho^2 D^2 \qquad (3)$$

where $G$ is the gravitational constant. This relation follows from the hydrostatic equilibrium equation. The material density $\rho$ is assumed to be in the range of 1.3 g cm$^{-3}$ to 1.7 g cm$^{-3}$. The former value is the average for C-type asteroids (Scheeres et al., 2015; Chesley et al., 2014); the latter is the bulk density of CM carbonaceous chondrites (e. g., Cold Bokkeveld) at an average porosity of ~25% (Consolmagno et al., 2008; Opeil et al., 2010). The difference between these two values is assumed to be due to the differences in macroporosity and water abundance between these two kinds of objects.

For the pressure in Ryugu's center from Eq. (3) and the above range of density values, we obtain $p$ = 45–76 Pa. This value is far below that for the triple point of water, equal to 610 Pa, therefore liquid water cannot exist in the Ryugu's interiors. At the vapor pressure of ice in the above range $p$ = 45–76 Pa the sublimation–condensation temperature of water ice is $T_w$ = 245–250 K that is a little bit lower than or near the asteroid mean temperature $T$ = 250–260 K. This relation of temperatures does not allow the ice to remain in the Ryugu's center for too long period of time. If the permeability of the asteroid's material is sufficiently small, however, some portions of water vapor could remain in closed pores in the center of Riugu.

However, if Ryugu has moved to its present orbit rather recently, it might not have heated enough to partially or completely lose ice, the initial stocks of which correspond to the abundance of water ice in the main-belt C-complex and similar asteroids (Schorghofer, 2016).

That is why it is of interest to estimate the characteristic timescale for heating the asteroid's interiors, which is defined by relation

$$t_h \sim \frac{D^2}{12\kappa} = \frac{1}{12}\left(\frac{D\rho C}{\Gamma}\right)^2, \qquad (4)$$

where $D \approx 870$ m is the diameter of Ryugu, $\kappa$ is its thermal diffusivity related to thermal inertia $\Gamma$, density $\rho$, and specific heat capacity $C$ by the equation $\Gamma = \rho C \sqrt{\kappa}$.

The heat capacity is estimated as follows. As Ryugu is classified as a Cg type (Binzel et al., 2001, 2002), its material is thought to be carbonaceous chondritic and corresponds in spectral characteristics to CM chondrites (e.g., Cloutis et al., 2011). For this case, we adopt the widely used analytic fit of Yomogida and Matsui (1984) for heat capacity of the non-differentiated bulk chondritic material

$$C(T) = 800 + 0.25T - 1.5 \times 10^7 \, T^{-2} \qquad (5)$$

which is a good fit for the heat capacity of the CM2 chondrite Cold Bokkeveld (Opeil et al., 2010) as well as achondrite meteorite shergottite Los Angeles (Opeil et al., 2012). For estimation of the value of $C(T)$ we accept a steady-state value of temperature below the asteroid's seasonal skin layer as $T = 250$ K, which is about the equilibrium Ryugu's temperature. At $T = 250$ K, Eq. (5) gives $C \approx 620$ J kg$^{-1}$ K$^{-1}$. For the above values of the parameters used in Eqs. (3) and (4) with the adopted span of density $\rho=1.3$–$1.7$ g cm$^{-3}$, we obtain the heating timescale in the range $t_h = (1.5$–$9.8)\times 10^4$ yr. Thus, for these basic average parameter values $\rho = 1500$ kg m$^{-3}$, $\Gamma = 200$ J K$^{-1}$m$^{-2}$ s$^{-1/2}$, and the above value of heat capacity we have $t_h \approx 4\times 10^4$ yr. If Ryugu reached its current orbit more recently (less than $4\times 10^4$ yr ago), it would retain a significant amount of its primordial abundance of water ice. It is important to note the period of volatile exhaustion (due to its dissipation) should correspond (if present) to the dynamical lifetime of the body in the near-Earth space.

If some amount of water ice is preserved in the interiors of Ryugu in spite of proceeding of ice sublimation, the water vapor could be delivered to the bottom of the subsurface layer, possibly by the use of existing cracks and microcracks. The survival of water ice in the upper subsurface layers is less probable. A thermal mechanism of short-lived outbursts due to the formation of deeper cracks and outgassing super volatile compounds was proposed in the course of the recent study of the comet 67P/Churyumov-Gerasimenko (Skorov et al., 2016). From the above consideration, we assume that because of the presence of the (ice-free) subsurface layer, the maximum subsolar temperature on the NEA at the moment of perihelion passage does not reach the deep, water-containing (in the form of vapor or ice) layers. Therefore one could expect a time delay (or "thermal lag") between the moment of maximum heating of the asteroid surface and that of the upper boundary of the water-containing reservoir.

In this case, the delay between perihelion passage by Ryugu and the onset of its suggested sublimation activity should include a time interval for the heat transport from the surface to an internal water vapor/ice reservoir ($t_1$) and a time for transfer of H$_2$O vapor from *this reservoir* to the surface ($t_2$). Because of the relatively small size of the asteroid and the likely permeable structure of the upper (regolith) layer, we assume $t_1 \gg t_2$ and neglect the value $t_2$. We take into account (derived from spectroscopic data) the nearly symmetrical character of Ryugu's possible sublimation process in time: nearly a month before aphelion passage and a month after the passage (Table 1). Based on these data, we assume the delay of sublimation from perihelion passage is roughly equal to a half of its orbital period $P$ (Fig. 1, spectrum on 11 July 2007; Fig. 2, the moment of sublimation activity is marked by a large star). As follows from these and other observations, after a relatively short interval of ejection of water vapor (and possibly fine dust) Ryugu becomes inactive again, up to its next passage of aphelion (Fig. 1, spectrum on 11/10 September 2007; Fig. 2, a small star). Thus, we calculate the delay through the phase lag of the seasonal temperature oscillations. The phase lag forms when the temperature oscillations from the surface penetrate through the thermal skin layer and reach a possible water vapor or ice (less probable) reservoir. The amplitude of these oscillations $\Delta T$ decreases with depth $z$ exponentially (Turcotte and Schubert, 2002)

$$\Delta T(z) = \Delta T_s \exp(-z/d), \tag{6}$$

where $\Delta T_s$ is the amplitude of seasonal temperature oscillations on the asteroid's surface, $d$ is the thickness of the thermal skin layer. As follows from Eq. (6), at the bottom of the layer, the amplitude of the seasonal temperature oscillations decreases by a factor of $e$. The value of the skin-layer thickness $d$ is

$$d = \sqrt{\frac{P\kappa}{\pi}} = \frac{\Gamma}{\rho C}\sqrt{\frac{P}{\pi}}, \tag{7}$$

where $P$ is the orbital period which for Ryugu equals 1.3 yr. Other parameters in Eq. (6) and their adopted values are explained above, after Eq. (4). The phase lag at depth $z$ is (Turcotte and Schubert, 2002) $\varphi = z / d$. In the case considered (at $t = \frac{1}{2} P$), we get $\varphi = 2 \pi t / P = \pi$. This relation can be used to obtain the depth of the level that is reached by the thermal wave during a half of the orbital period. The depth is equal to $z_1 = \pi d$, where the relation for $d$ is given by Eq. (7). The amplitude of temperature variations at this depth is given by Eq. (6) as $\Delta T(\pi d) = \Delta T_s \exp(-\pi)$. For the case of Ryugu, we obtain the thickness of the thermal skin layer $d = 0.53–1.4$ m and $z_1 = 1.7–4.4$ m, with the most probable values $d \approx 0.8$ m and $z_1 \approx 2.5$ m, respectively. These values correspond to the accepted above average values of parameters $\rho$ and $\Gamma$. The amplitude of seasonal subsolar (equatorial) temperature cycle $\Delta T_s \approx 60$ K (Fig. 4) gives the half-period temperature variations $\Delta T(\pi d) \approx 3$ K at $z = z_1$ (the level which is reached by the thermal wave after the half orbital period). This would suggest that at the depth about 2–4 meters some reservoir of water vapor or ice exists, ice being less probable. The reservoir is activated by the seasonal thermal wave with an amplitude of several kelvins.

Thus, the presumed temporal sublimation/degassing activity of Ryugu indicates a not very long period of the asteroid's staying in the current orbit. This implies a relatively recent transition of the asteroid from the main belt to the near-Earth area. Figure 5 shows qualitatively the proposed internal structure for Ryugu and suggested mechanism of water ice sublimation.

Another potential cause of sublimation on Ryugu can also be proposed. A similar observational pattern of differences in reflectance spectra could originate in the case of residual ice present on the Ryugu's surface near one of its poles due to a considerable change of insolation conditions. According to recent estimations of spin-vector orientation derived from a combined analysis of data of visual lightcurves and mid-infrared photometry and spectroscopy (Müller et al., 2017), Ryugu has a retrograde rotation with the most likely axis orientation of ($\lambda$, $\beta)_{ecl}$ = (310°-340°, -40°± ~ 15°), a rotation period of $P_{sid} = 7.63109^h$, and a very low surface roughness (r. m. s. of surface slopes < 0.1). Then, one could imagine that, at these mentioned asteroid parameters, some ice-containing material preserved in a surface depression (a cold trap) near the North Pole of the asteroid would be more sunlit near aphelion then at perihelion. Additional heating of the icy material could produce the observed sublimation activity of Ryugu near aphelion. However, owing to the possibility of a relatively quick change of an NEA's orbital parameters under the action of gravitational perturbations from terrestrial planets, the latter mechanism could represent a temporal phenomenon implying that the residence time of the asteroid in the near-Earth space is not very long in order for it to retain the icy matter near the surface. This is so especially when taking into account the significant obliquity of Ryugu's rotational axis (~60°) (Müller et al., 2017).

Besides, as thermogravimetric measurements (Frost et al., 2000) show, phyllosilicates (being the main constituents of the matrix of carbonaceous chondrites) have three stages of dehydration with temperature elevation. Such compounds as ferruginous smectite and some nontronites lose a significant proportion of mass at 300–360 K (12–17%), at 360–380 K (3–4%), and at 380–430 K (1–2%) (Frost et al., 2000). The first two temperature intervals correspond well to the calculated range of subsolar temperatures of Ryugu ~326- to- 394 K (Fig. 4). So, the first two stages of dehydration process in phyllosilicates may be an additional source of Ryugu's surface/subsurface activity.

Nevertheless, both the above mechanisms (depletion of internal water ice and sublimation of near-polar ice deposits and/or dehydration of phyllosilicates) are possible and could give inputs to sublimation/outgassing and dust activity of Ryugu near aphelion.

### 4. Discussion

Due to early aqueous differentiation of volatile-abundant matter in primitive asteroids or their parent bodies because of the decay of short-lived radionuclides ($^{26}$Al, $^{60}$Fe, etc.) (e. g., Srinivasan et al., 1999; Ghosh et al., 2006; Wadhwa et al., 2006), sufficiently large bodies accumulated ices in the subsurface layers due to differentiation and formation of an ice mantle (Grimm and McSween, 1993; Busarev et al., 2003, 2005). The calculated subsolar temperature ranges of the main-belt asteroids mentioned above are below the melting point of water ice (Busarev et al., 2015, 2016) but they are higher than the threshold temperature for $H_2O$ ice sublimation (~145 K) (Schorghofer, 2008). This means that the majority of primitive main-belt asteroids predominantly slowly and gradually deplete their ice stock in the process of sublimation, this process being faster at shorter heliocentric distances (e. g., Schorghofer, 2008). There are not numerous bodies that rather effectively lose their ice like (1474) Beira: because of the large eccentricities of their orbits, they experience much stronger heating by solar radiation near perihelion and, hence, faster exhaustion of the subsurface ices (Busarev et al., 2016). Besides, regolith on primitive main-belt asteroids has likely been formed in the usual way (due to repeated impacts of meteoroids), but should also contain ice inclusions if some ice layers existed in the asteroids' interiors. On the other hand, main-belt asteroids should have an ancient surface on average. The common features of these ancient surfaces are likely a considerable thickness of the regolith layer and its very low heat conductivity, similar to those of the lunar regolith layer. These features lead to damping of a diurnal thermal wave in a surface layer of only several centimeters (e. g., Wechsler et al., 1972; Langset and Keihm, 1975; Kuzmin and Zabalueva, 2003). Thus, we could imagine on a qualitative level the internal structure of primitive main-belt asteroids as a combination of phyllosilicate and ice layers (and/or inclusions) with a thick regolith envelope. In any case, sublimation activity on a primitive asteroid becomes likely more intense (not only at perihelion) after excavation of fresh ice by recent meteoroid impact(s).

A main-belt primitive asteroid can break apart and form NEAs due to a strong impact event or the action of gravitational resonances. Then the new NEAs quickly (after several revolutions around the Sun) lose a predominant part of the ices stored near their surfaces because of the considerable growth of their subsolar and average subsurface temperatures by ~100 degrees (e. g., Delbo, Michel, 2011). Tidal interactions with the Earth and other terrestrial planets could raise fractures and breaks in a NEA body, that facilitate venting of any volatiles from its interior and their loss from the uppermost regolith. A similar effect, but only in a thin layer, can be produced by the sharp temperature differences between day and night sides of a NEA with rotation on a timescale of several hours. As follows from model calculations, the depth of a "desiccated" layer depends on amplitude of the surface diurnal and seasonal temperature variations and characteristics of asteroid regolith and subsurface matter such as the density, structure, porosity, heat conductivity, and so on (e. g., Delbo, Michel, 2011; Skorov et al., 2016, and this paper). Thus, because of the presence of the desiccated layer, the maximum subsolar temperature on a NEA at perihelion passage still does not reach its frozen interiors. In other words, we expect a time delay (or "thermal lag") between the time of maximum heating the asteroid surface and that of the outer boundary of an internal icy reservoir. Moreover, during each passage of perihelion by the asteroid, the outer boundary of the ice core should move deeper. Importantly, a destructive action of the seasonal "heat wave" (reaching 50-100 degrees on the surface and several degrees at the bottom of the thermal skin layer) onto the asteroid matter including closed ice traps represents likely a mechanism of the formation of numerous new microcracks and, as a result, elevation of permeability of the matter for volatiles. A similar thermal mechanism of short-lived outbursts because of formation of deeper cracks and outgassing super volatile compounds was proposed under recent investigations of 67P/Churyumov-Gerasimenko comet (Skorov et al., 2016).

One more important issue deserves attention. One of the key characteristics of an asteroid is the degree of its shock and impact processing, due to the majority of small planets originated as a result of the crushing of larger parent bodies. This is reflected in the irregular shape of almost all main-belt asteroids with defined light curves (more than 5500 asteroids) according to the Minor Planet Center (http://alcdef.org). Another consequence of this process must be the degree of change of the internal structure in each body. The same could be true for NEAs. Despite widespread notions that most NEAs are reaccumulated rubble-piles (e. g., Bottke et al., 2005; Binzel, Kofman, 2005; Goldreich, Sari, 2009), our model of Ryugu (Fig. 5) corresponds to a cracked monolithic body. Such a structure allows venting volatiles (but not as fast as for a rubble pile). The model used in this paper is in agreement with observations confirming the existence of monolithic asteroids among NEAs. The obvious examples are fast-rotating asteroids (e. g., Asphaug, Scheeres, 1999; Whiteley et al., 2002). We prefer this model for Ryugu because rubble reaccumulation should considerably decrease the lifetime of any volatiles on a primitive NEA and therefore reduce the probability of detection of their signs.

A simpler explanation is also possible. A similar observational pattern could have its origin in the case of residual ice present at one of Ryugu's poles, due to a considerable change of lighting (insolation) conditions. According to recent estimations of spin-vector orientation derived from the combined analysis of data sets of visual lightcurves and mid-infrared photometry and spectroscopy (Müller et al., 2017), Ryugu has a retrograde rotation with the most likely axis orientation of $(\lambda, \beta)_{ecl}$ = (340°, -40°), a rotation period of $P_{sid}$ = 7.63109$^h$, and a very low surface roughness (r.m.s. of surface slopes < 0.1). Thus, one could imagine that, at the mentioned asteroid orientation parameters, some ice-content material preserved in a surface depression at the North Pole of the asteroid would be more sunlit near aphelion then at perihelion. Additional heating of the icy material could produce the observed sublimation activity of Ryugu in the vicinity of aphelion. The simplicity of this explanation renders it to look more realistic. But, it should be emphasized that owing to a relatively quick change of an NEA's orbital parameters under the action of gravitational perturbations from terrestrial planets, ice preservation in this way is likely not a long-term phenomenon. This suggests that the residence time of the asteroid in the near-Earth space should not be very long in order to keep the icy matter at the surface.

Obviously, the direct observational confirmation of the described ideas on the mechanism of sublimation / dust activity of the NEA is only the discussed reflectance spectrum of Ryugu obtained by Vilas (2008) close to the aphelion of the asteroid. However, it seems there is additional important indirect evidence, namely detection of significant spectral variations on a NEA with similar characteristics, (101955) Bennu. Although such changes on Bennu have been registered only in the infrared range around aphelion, their sharp character (change of the gradient of the reflectance spectrum from negative to positive) point to possible spectral variations in the short-wavelength part of asteroid spectrum too and, hence, confirm the same transient process of sublimation activity as on Ryugu. Binzel et al. (2015) have another opinion about nature of the discussed spectral variation on Bennu (very different sizes of particles forming an equatorial ridge and polar regions of the asteroid) but we propose that our explanation is more feasible.

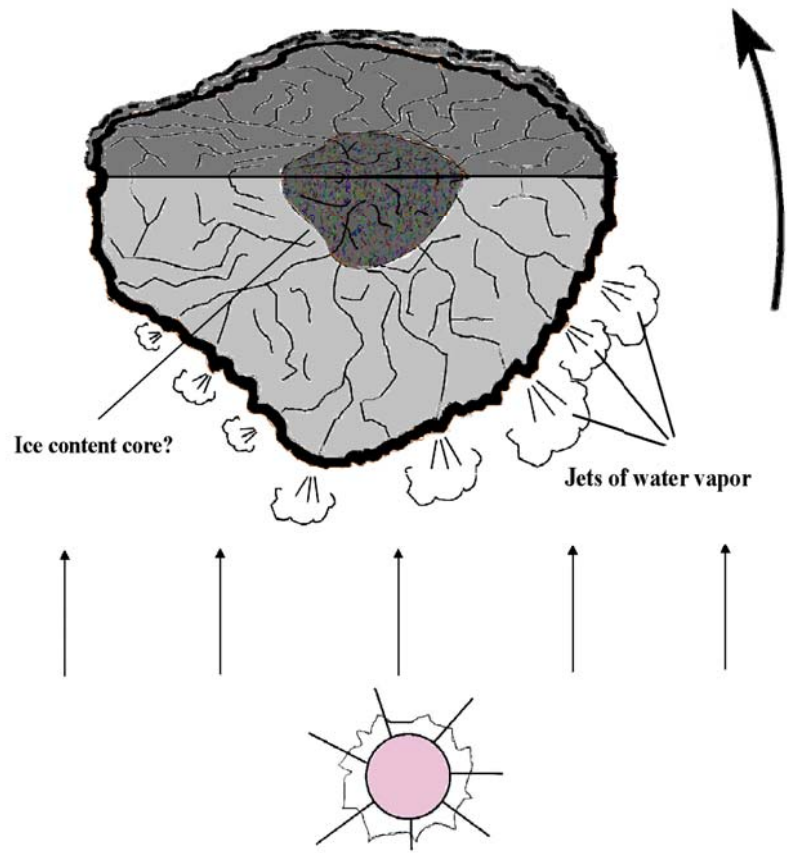

Fig. 6. Suggested internal structure of a primitive-type NEA including an icy core and possible mechanism of its sublimation/ dust activity near aphelion. Large curved arrow shows direction of the asteroid rotation.

**5. Conclusions**

The orbital ellipticity of a small planet trajectory makes it possible to test the volatility of the body's material depending on heliocentric distance and the accompanying surface temperature changes. The above results show that any main-belt asteroids that accumulated ices beneath the surface over their lifetime could demonstrate such activity near perihelion. However, the phenomenon on NEAs has important differences: in the short time of an NEAs evolution, it begins to work in "a reverse order". Critical consideration of available spectroscopic data on (162173) Ryugu and creation of a preliminary thermophysical model of this and similar primitive asteroids allowed us (Busarev et al., 2017b and this paper) for the first time demonstrate the possibility of a sublimation process on such bodies in unexpected order – at the passage of aphelion.

The presumed temporal sublimation/degassing activity of (162173) Ryugu is likely an evidence of existence of a residual frozen core in its interiors. Any sublimation/dust activity of Ryugu, and the possible comparatively small depth of its putative ice reservoir, could indicate a relatively short residence time of the asteroid in the near-Earth space. A similar observational pattern could originate in the case of residual ice presence at one of Ryugu's poles due to a considerable change of lighting conditions. Some additional heating of the icy material could produce the observed sublimation activity of Ryugu near aphelion. But, owing to the relatively quick change of an NEA's orbital parameters under the action of gravitational perturbations from terrestrial planets, this is likely only a short-term phenomenon.

Presently, our knowledge about Ryugu is far from complete. We cannot yet decide which of the noisy spectral data are bad or correct. To verify the described suggestions, we need new more thorough ground-based spectral rotationally-resolved observations of Ryugu and other NEAs of primitive types and, finally, arrival of *Hayabusa 2* to the asteroid.

*Acknowledgements.* The authors are grateful to Daniela Lazzaro and Andreas Nathues for providing spectral data on Ryugu and Ceres, respectively. The authors thank anonymous referees for useful comments.

**References:**

Abe, M., Kitazato, K., Sarugaku, Y., Kawakatsu, Y., Kinoshita, D., 2007. Ground-based observation of post-Hayabusa mission targets. 38$^{th}$ Lunar and Planet. Sci. Conf., Abstract No. 1638.

A'Hearn, M. F., Feldman, P. D., 1992. Water vaporization on Ceres. Icarus 98, 54-60.

Asphaug, E., Scheeres, D. J., 1999. Deconstructing Castalia: Evaluating a postimpact state. Icarus, 139, 383-386

Bottke, W. F., Durda, D. D., Nesvorný, D., Jedicke, R., Levison, H. F., 2005. Linking the collisional history of the main asteroid belt to its dynamical excitation and depletion. Icarus 179, 63-94.

Binzel, R. P., Harris, A. W., Bus, S. J., Burbine, T. H., 2001. Spectral properties of near-Earth objects: Palomar and IRTF results for 48 objects including spacecraft targets (9969) Braille and (10302) 1989 ML. Icarus 151, 139-149.

Binzel, R. P., Kofman, W., 2005. Internal structure of Near-Earth Objects. C. R. Physique 6, 321-326.

Binzel, R. P., Lupishko, D. F., Di Martino, M., Whiteley, R. J., Hahn, G. J., 2002. Physical properties of Near-Earth objects. In: Asteroids III (W. F. Bottke, A. Cellino, P. Paolicchi, and R. P. Binzel, eds.), Univ. of Arizona Press, pp. 255-271.

Binzel, R. P., DeMeo, F. E., Burt, B. J. and 14 co-authorths, 2015. Spectral slope variations for OSIRIS-REx target Asteroid (101955) Bennu: Possible evidence for a fine-grained regolith equatorial ridge. Icarus 256, 22-29.

Britt, D. T., Yeomans, D., Housen, K., Consolmagno, G., 2002. Asteroid Density, Porosity, and Structure. In: Asteroids III (W. F. Bottke, A. Cellino, P. Paolicchi, and R. P. Binzel, eds.), Univ. of Arizona Press, 485-500.

Busarev, V. V., Dorofeeva, V. A., Makalkin, A. B., 2003. Hydrated silicates on Edgeworth-Kuiper objects – probable ways of formation. Earth, Moon and Planets 92, 345-357.

Busarev, V. V., Dorofeeva, V. A., Makalkin, A. B. Possibility of separating silicates and organics in large Kuiper belt objects. Lunar & Planetary Science Conf. 36th, Houston, 2005, Abs. #1074.

Busarev, V. V., Barabanov, S. I., Rusakov, V. S., Puzin, V. B., Kravtsov, V. V., 2015. Spectrophotometry of (32) Pomona, (145) Adeona, (704) Interamnia, (779) Nina, (330825) 2008 XE3, and 2012 QG42 and laboratory study of possible analog samples. Icarus 262, 44-57.

Busarev, V.V., Barabanov, S.I., Puzin, V.B., 2016. Material composition assessment and discovering sublimation activity on asteroids 145 Adeona, 704 Interamnia, 779 Nina, and 1474 Beira. Solar System Research 50, 281–293.

Busarev, V. V., Barabanov, S. I., Scherbina, M. P., Puzin, V. B., 2017a. Sublimation activity of (145) Adeona, (704) Interamnia, (779) Nina, and (1474) Beira and some confirmations. 48th Lunar and Planet. Sci. Conf., Woodlands, Texas, Abstract No. 1919.

Busarev, V. V., Makalkin, A. B., Vilas, F., Barabanov, S. I., Scherbina, M. P. 2017b. New candidates for active asteroids: Main-belt (145) Adeona, (704) Interamnia, (779) Nina, (1474) Beira, and near-Earth (162173) Ryugu. Icarus. DOI: 10.1016/j.icarus.2017.06.032.

Campins, H., de León, J., Morbidelli, A., Licandro, J., Gayon-Markt, J., Delbo, M., Michel, P., 2013. The origin of asteroid 162173 (1999 JU3). Astron. J. 146, 26-31.

Chesley, S. R., Farnocchia, D., Nolan, M. C. and 13 co-authors, 2014. Orbit and bulk density of the OSIRIS-REx target Asteroid (101955) Bennu. Icarus, 235, 5-22.

Clark, B. E., Binzel, R. P., Howell, E. S. and 12 co-authors, 2011. Asteroid (101955) 1999 RQ36: Spectroscopy from 0.4 to 2.4 μm and meteorite analogs. Icarus 216, 462-475.

Cloutis, E. A., Hudon, P., Hiroi, T., Gaffey, M.J., Mann, P., 2011. Spectral reflectance properties of carbonaceous chondrites: 2. CM chondrites. Icarus 216, 309-346.

Consolmagno, G.J., Britt, D.T., Macke, R.J., 2008. The significance of meteorite density and porosity. Chem. Erde 68, 1-29.

Delbo, M., Michel, P., 2011. Temperature history and dynamical evolution of (101955) 1999 RQ 36: A potential target for sample return from a primitive asteroid. Astroph. J. Lett., 728, L42-L6.

Frost, R., Ruan, H., Kloprogge, J. T., Gates, W., 2000, Dehydration and dehydroxylation of nontronites and ferruginous smectite. Thermochim. Acta 346, 63-72.

Hansen, J. E., Travis, L. D., 1974. Light scattering in planetary atmosphere. Space Sci. Rev. 16, 527-610.

Gaffey, M. J., Burbine, T. H., Binzel, R. P., 1993. Asteroid spectroscopy and Progress and perspectives. Meteoritics 28, 161-187.

Ghosh, A., Weidenschilling, S. J., McSween Jr., H. Y., Rubin, A., 2006. Asteroidal heating and thermal stratification of the asteroid belt. In: Meteorites and the early solar system II (D. S. Lauretta and H. Y. McSween, Jr., eds.), Univ. of Arizona Press, pp. 555-566.

Goldreich, P., Sari, R., 2009. Tidal evolution of rubble piles. Astrophys. J. 691, 54-60.

Grimm, R. E, McSween Jr., H. Y., 1993. Heliocentric zoning of the asteroid belt by aluminum-26 heating. Science 259, 653-655.

Ishiguro, M., Kuroda, D., Hasegawa, S. and 31 co-authors, 2014. Optical properties of (162173) 1999 JU3: In preparation for the Jaxa *Hayabusa 2* sample return mission. Astrophys. J. 792, 74-82.

Küppers, M., O'Rourke, L., Bockelée-Morvan, D., and 10 co-authors, 2014. Localized sources of water vapour on the dwarf planet (1) Ceres. Nature 505, 525-527.

Kuzmin, R. O., Zabalueva, E. V., 2003. The temperature regime of the surface layer of the Phobos regolith in the region of the potential *Fobos–Grunt* space station landing site. Solar Sys. Res. 37, 480-488.

Langset, M. S., Jr., Keihm, S. J., 1975. Direct measurements of the thermal flux on the Moon. In: Cosmochemistry of the Moon and planets, Proc. Soviet–American Conf. on Cosmochemistry of the Moon and Planets, Moscow: Nauka, pp. 200–209 (in Russian).

Lazzaro, D., Barucci, M. A., Perna, D., Jasmim, F. L., Yoshikawa, M., Carvano, J. M. F., 2013. Rotational spectra of (162173) 1999 JU3, the target of the Hayabusa 2 mission. Astron. Astrophys. 549, 2-5.

Li, J.-Y., Helfenstein, P., Buratti B. J., Takir D., Clark B. E., 2015. Asteroid Photometry. In: Asteroids IV (P. Michel et al., eds.), Univ. of Arizona Press, pp. 129-150.

Lissauer, J. J., de Pater, I., 2013. Fundamental planetary science: physics, planetary and habitability. Cambridge Univ. Press, 583 pp.

McSween, H. Y., Jr., Lauretta, D. S., Leshin, L. A., 2006. Recent advances in meteoritics and cosmochemistry. In: Meteorites and the Early Solar System II (D. S. Lauretta and H. Y. McSween, Jr., eds.), Tucson, AZ: Univ. Arizona Press, pp. 53-66.

Mie, G., 1908. Beiträge zur Optik trüber Medien speziell kolloidaler Goldlösungen. Ann. Phys. 25, 377–445.

Moskovitz, N. A., Abe, S., Pan, K.-S., and 9 co-authors, 2013. Rotational characterization of Hayabusa II target Asteroid (162173) 1999 JU3. Icarus 224, 24-31.

Müller, T. G., Ďurech, J., Ishiguro, M., and 27 co-authors, 2017. Hayabusa-2 Mission Target Asteroid 162173 Ryugu (1999 JU3): Searching for the Object's Spin-Axis Orientation. Astron. & Astrophys. 599, 103–127.

Nathues, A., Hoffmann, M., Schaefer, M., and 25 co-authors, 2015. Sublimation in bright sports on (1) Ceres. Nature 528, 237-240.

Opeil, C. P., Consolmagno, G. J., Britt, D. T., 2010. The thermal conductivity of meteorites: New measurements and analysis. Icarus 208, 449−454.

Opeil, C. P., Consolmagno, G. J., Safaric, D. J., Britt, D. T., 2012. Stony meteorite thermal properties and their relationship with meteorite chemical and physical states. Met. Planet. Sci. 47, 319-329.

Schorghofer, N., 2016. Predictions of depth-to-ice on asteroids based on an asynchronous model of temperature, impact stirring, and ice loss. Icarus 276, 88-95.

Scheeres, D. J., Britt, D., Carry, B., Holsapple, K. A., 2015. Asteroid interiors and morphology. In: Asteroids IV (P. Michel, F. E. DeMeo, and W. F. Bottke, eds.), Univ. of Arizona Press, Tucson, pp. 745–766.

Skorov, Yu. V., Rezac, L., Hartogh, P., Bazilevsky, A. T., Keller, H. U., 2016. A model of short-lived outbursts on the 67P/CG from fractured terrains. Astron. Astrophys. 593, 76-85.

Srinivasan, G., Goswami, J. N., Bhandari, N., 1999. $^{26}$Al in eucrite Piplia Kalan: Plausible heat source and formation chronology. Science 284, 1348-1350.

Sugita, S., Kuroda, D., Kameda, S., Hasegawa, S., Kamata, S., Hiroi, T., Abe, M., Ishiguro, M., Takato, N., Yoshikawa, M., 2013. Visible spectroscopic observations of asteroid 162173 (1999 JU$_3$) with the Gemini-S telescope. 44$^{th}$ Lunar Planet. Sci. Conf., Abs. No. 2591.

Turcotte, D. L., Schubert G., 2002. Geodynamics. Cambridge University Press (second edition), Cambridge, England, 456 p.

Tholen, D. J., 1989. Asteroid taxonomic classifications. In: Asteroids II (Binzel R. P., Gehrels T. and Mattews M. S., eds.), Tucson: Univ. of Arizona Press, pp. 1139-1150.

Vilas, F., Jarvis, K. S., Gaffey, M. J., 1994. Iron alteration minerals in the visible and near-infrared spectra of low-albedo asteroids. Icarus 109, 274-283.

Vilas, F., 2008. Spectral characteristics of *Hayabusa 2* Near-Earth asteroid targets 162173 1999 JU3 and 2001 QC34. Astron. J. 135, 1101-1105.

Vilas, F., 2012. New Spectral Reflectance Observations of Hayabusa 2 Near-Earth Asteroid Target 162173 1999 JU3. American Astronomical Society DPS meeting # 44, 102.03.

Wadhwa, M., Srinivasan, G., Carison, R. W., 2006. Timescales of planetesimal differentiation in the early solar system. In: Meteorites and the early solar system II (D. S. Lauretta and H. Y. Jr. McSween, eds.), Univ. of Arizona Press, 715-731.

Wechsler, A.E., Glaser, P.E., Fountain, J.A., 1972. Thermal Properties of Granulated Materials. In: Thermal Characteristics of the Moon (J.W. Lucas, ed.), Cambridge: MIT Press, pp. 215–241.

Whiteley, R. J., Hergenrother, C. W., Tholen, D. J., 2002. Monolithic fast-rotating asteroids. Proc. of ACM 2002, Tech. Univ. Berlin, Berlin, Germany (ESA-SP-500), pp. 473-480.

Yomogida, K., Matsui, T., 1984. Multiple parent bodies of ordinary chondrites. Earth Planet. Sci. Lett. 68, 34-42.

Yu, L. L., Ji, J. H., 2015. Surface thermophysical properties determination of *OSIRIS-REx* target asteroid (101955) Bennu. MNRAS 452, 368−375.